\documentclass[aps,prl,twocolumn,groupedaddress]{revtex4}
\usepackage{graphicx}
\usepackage{amsmath}
\usepackage{epstopdf}

\begin{document}


\title{Self Assembly of Soft Matter Quasicrystals and Their Approximants}

\author{Christopher R. Iacovella$^{1,3,\dagger}$}
\author{Aaron S. Keys$^{1,\dagger}$}
\author{Sharon C. Glotzer$^{1,2}$}
\affiliation{$^1$ Department of Chemical Engineering and $^2$Department of Materials Science and Engineering \\University of Michigan, Ann Arbor, Michigan 48109-2136\\
$^3$ Department of Chemical and Biomolecular Engineering, Vanderbilt University, Nashville, Tennessee 37235-1604 (Present address) \\
$^\dagger$These authors contributed equally to this work }

\date{\today}

\begin{abstract}

The surprising recent discoveries of quasicrystals and their approximants in soft matter systems poses the intriguing possibility that these structures can be realized in a broad range of nano- and micro-scale assemblies. It has been theorized that soft matter quasicrystals and approximants are largely entropically stabilized, but the thermodynamic mechanism underlying their formation remains elusive.  Here, we use computer simulation and free energy calculations to demonstrate a simple design heuristic for assembling quasicrystals and approximants in soft matter systems.  Our study builds on previous simulation studies of the self-assembly of dodecagonal quasicrystals and approximants in minimal systems of spherical particles with complex, highly-specific interaction potentials.  We demonstrate an alternative entropy-based approach for assembling dodecagonal quasicrystals and approximants based solely on particle functionalization and shape, thereby recasting the interaction-potential-based assembly strategy in terms of simpler-to-achieve bonded and excluded-volume interactions.  Here, spherical building blocks are functionalized with mobile surface entities to encourage the formation of structures with low surface contact area, including non-close-packed and polytetrahedral structures. The building blocks also possess shape polydispersity, where a subset of the building blocks deviate from the ideal spherical shape, discouraging the formation of close-packed crystals.  We show that three different model systems with both of these features -- mobile surface entities and shape polydispersity -- consistently assemble quasicrystals and/or approximants.  We argue that this design strategy can be widely exploited to assemble quasicrystals and approximants on the nano- and micro- scales.  In addition, our results further elucidate the formation of soft matter quasicrystals in experiment.  

\end{abstract}
\maketitle

Until fairly recently, quasicrystals and their approximants have been observed only in atomistic systems.  Over the past decade, there have been sporadic reports of quasicrystals and approximants in nanometer and micron-scale systems.  Examples include holographically-trapped~\cite{grier} and laser-field-induced~\cite{mikhael, glotzer08} quasicrystalline materials made of micron-sized spheres, self-assembled quasicrystals and approximants formed by spherical dendrimer micelles~\cite{ungar03, zeng04}, phase-separated star-triblock copolymers~\cite{dotera}, binary nanoparticle superlattices~\cite{talapin2009}, spherical micelles of phase-separated block copolymers~\cite{lee10, fischer11}, and simulations of hard tetrahedra~\cite{amir09}.  These reports pose an intriguing possibility that these structures might be assembled in a broad range of systems.   In one such system, spherical dendrimeric micelles functionalized with alkyl tails form a dodecagonal (12-fold) quasicrystal (DQC), as well as other non-close packed structures such as the body-centered cubic (bcc) and A15 crystals~\cite{balagurusamy_jacs}.  In similar systems, various types of block copolymer micelles arrange into quasicrystals with 12-fold, and possibly 18-fold, symmetry~\cite{fischer11}, as well as various periodic approximants~\cite{lee10}.  

The dendrimer and block copolymer micelle systems in particular all share an important common feature: their constituent micelles exhibit a soft ``squishy corona'' in which terminal groups avoid each other to minimize steric interactions. It has been predicted that this mechanism causes the system to adopt structures that minimize surface contact area between neighboring micelles~\cite{kamien,kamien00}.  The structure that minimizes surface contact area, known as the Weaire-Phelan or A15 structure~\cite{weaire}, is structurally similar to a DQC, but, since DQCs do not minimize surface contact area, other factors must contribute to their stability.  It has been suggested that entanglement of terminal groups may give rise to three-body entropic effects that favor quasicrystals in systems of monodisperse micelles~\cite{lifshitz2007soft, barkan2010stability}.  In all these micellar systems, entropic effects appear to play a predominant role in stabilizing the quasicrystals and approximants, potentially distinguishing them from many of their atomistic counterparts in which strong attractive interactions are present.

Computer simulation studies of self-assembly have demonstrated that quasicrystals can be assembled by an inverse-design mechanism.  In particular, pair potentials can be designed to make close-packing unfavorable, causing such systems to instead form quasicrystals and approximants ~\cite{dzug93, roth00, engel08}.  These complex interaction potentials have yet to be realized in experimental systems on the micro- or nano-scale, but we propose that a similar effect can be achieved via shape polydispersity, where a subset of the micelles deviate from the ideal spherical shape.  Shape polydispersity arises naturally in many micelle-forming systems, and, in general, particle shape is a tunable parameter in many micro- and nano-scale systems~\cite{glotzer2007}.  

In this article, we introduce a design strategy based on the ideas described above to direct the self-assembly of three-dimensional DQCs and/or their periodic approximants in systems of (approximately) spherical micelles or similarly shaped particles.  We study different types of nano/microscale building blocks with features that promote structures with low surface contact area and suppress close-packing.  Structures with low surface contact area are promoted by functionalizing spherical building blocks with mobile entities connected to their surface, similar to functionalized spherical dendrimers~\cite{zeng04}.  Close-packing is suppressed by incorporating shape polydispersity into the system in the form of particle asphericity.  Both features are relatively common aspects of soft matter and related systems and should be achievable experimentally; a schematic of our strategy is shown in Fig. 1a.  Applying this strategy in computer simulations, we show three key results. (1) We verify the theoretical predictions that interactions between terminal coatings can drive the system to form surface-area minimizing structures~\cite{kamien,kamien00}.  (2) We demonstrate that shape polydispersity can be used to suppress the formation of close-packed structures.  (3) We show that three different simulated micellar systems that possess both of these characteristics reproducibly form DQCs and/or approximants.  These models -- a simplified model of a spherical micelle and two micelle-forming systems composed of tethered nanosphere building blocks~\cite{zhang03, iacovella05, iacovella09, iacovella09b} -- represent the only simulated micellar systems currently known to form 3d quasicrystals or approximants through self-assembly.  Because the models are closely related to experimental systems known to form DQCs and/or approximants~\cite{zeng04, ungar03, lee10, fischer11}, our results may provide pertinent insight regarding their formation.  In the future, the assembly strategy that we employ may serve as a heuristic for expanding the range of systems that assemble DQCs and approximants.

\begin{figure}[h]
\includegraphics[width=.4\textwidth]{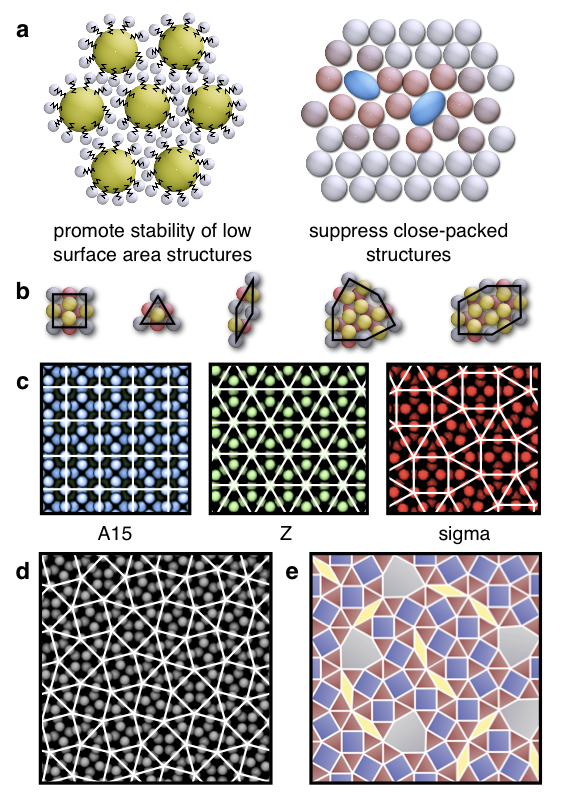}
\caption{Assembly strategy and structure of the dodecagonal quasicrystals and approximants.  (a) Schematic of the proposed two-part strategy that uses functionalization and shape to form DQCs.  Particle functionalization (left) promotes the formation of structures with low surface contact area and asphericity (right) inhibits the formation of close-packed structures.  Particles colored red in the asphericity schematic (right) are meant to highlight where the crystal is disrupted by the presence of aspherical particles (blue). (b) Valid tiles for the DQC. The DQC and approximants can be described as a periodic stacking of plane-filling arrangements of tiles in the z direction (out of the page).  The gray particles at the nodes of the tiles form layers at z=1/4 and z=3/4 and sit at the centers of 12-member rings.  The yellow particles and red particles form layers at z=0 and z=1/2 respectively.  In the DQC, the gray particles form a dodecagonal layer with 12-fold symmetry, and the yellow and red particles form hexagonal layers rotated by 30 degrees to obtain 12-fold symmetry.  (c) Three common DQC approximants. (d) A higher-order approximant generated through the inflation method (see text). (e) A representative DQC random tiling of squares, triangles, rhombs and shields, adapted from Ref. \cite{roth00epjb} }\label{fig:approx}
\end{figure}

\section{Dodecagonal quasicrystals and approximants}

We first introduce definitions and terminology that will facilitate our discussions in subsequent sections.  A crystal is defined as a structure with long-range positional order, as identified, for example, by the presence of Bragg peaks in the diffraction pattern~\cite{lifshitz2007crystal}.  A quasicrystal is a quasi-periodic crystal; that is, a crystal that lacks periodicity~\cite{lifshitz2003quasicrystals}, but still exhibits diffraction peaks.  Quasicrystals sometimes (but need not) exhibit rotational symmetries that are incompatible with periodicity.  Several types of quasicrystals have been observed in experiment, but in this article we focus on DQCs in particular because those are to date the most commonly reported type of quasicrystal in soft matter systems.  DQCs are characterized by their long-range dodecagonal (i.e, 12-fold) rotational symmetry.

DQCs are polytetrahedral structures~\cite{nelson1989polytetrahedral} of the Frank-Kasper (FK) type~\cite{frank59}.  For the class of FK structures considered here, ordered structures are distinguished by their ``tiling'' pattern, constructed by connecting the centers of neighboring 12-member rings of particles (see panels b-e of Fig.~1).  The structures are layered and, whether periodic or aperiodic in the plane, they repeat periodically in the direction orthogonal to the plane (into to the page in Fig.~1).    There are five valid tiles that can be arranged to form structures with complete 12-member rings without disorder.  These tiles take the shape of a square, a triangle, a rhomb, a shield, and an asymmetric hexagon~\cite{roth00}, and are illustrated in Fig.~1b.  Periodic arrangements of these tiles result in periodic crystals, sometimes known as ``approximants,'' that are indistinguishable from DQCs locally~\cite{goldman93}.  Three common approximants, known as the A15, Z, and sigma structures, are shown in Fig. 1c.  Increasingly complex approximants, such as the structure depicted in Fig.~1d, can be constructed by inflation, whereby tiles are sequentially replaced with smaller sub-tiles~\cite{janot97, zeng2006inflation}.  

In addition to periodic arrangements, non-periodic arrangements of tiles that fill the plane can also be constructed, resulting in quasicrystals.  Various methods can be used to construct the tilings; methods such as inflation~\cite{zeng2006inflation}, projection~\cite{janot97}, or matching rules~\cite{onoda1988growing, steinhardt1996simpler} produce deterministic quasicrystals, whereas random tilings~\cite{henley1991random} give rise to a range of similar quasicrystals with the tiles reshuffled locally, characterized by local phason fluctuations.  Imperfect quasicrystals of either type may also exhibit global phason strain whereby particular tiles or orientations of tiles occur more or less frequently that in the ideal case, giving rise to shifts and broadening of the diffraction peaks~\cite{lubensky1986distortion}.  Deterministic quasicrystals are thought to be energetically stabilized, whereas random tiling quasicrystals are thought to be entropically stabilized~\cite{henley1991random}.  Fig.~1e shows a typical random-tiling DQC~\cite{roth00epjb} that we envision might form in soft-matter systems, which are often stabilized by entropy.  The structure is composed mostly of squares and triangles, and is locally similar to the sigma approximant.   The sigma approximant is the thermodynamically stable state for many systems that form DQCs, and the two structures often arise in nearby regions of parameter space~\cite{ungar03, zeng04, talapin2009}.  The experimental protocol may dictate whether a metastable DQC or a stable sigma approximant is obtained.   In the case of the simulations we perform on model micelles, we are limited to relatively small, finite size simulations, as discussed subsequently. As such our systems are typically too small to unambiguously distinguish between quasicrystals and approximants, or identify phason strain.  With this caveat in mind, we refer to our assembled structures as quasicrystals if they are composed of valid tiles for the DQC, exhibit strong peaks in the diffraction pattern, and are not periodic (aside from the trivial periodicity imposed by the periodic boundary conditions on the scale of the sample).

\section{Simulation results}

We begin by performing  molecular dynamics simulations~\cite{lammps} of a simplified model of a spherical micelle (MSM) that considers only excluded volume interactions between terminal groups on the micelle surface (Fig. 2a).  Unlike the truly minimal ``fuzzy sphere'' micelle model of Ref.~\cite{kamien} that treats inter-micelle interactions with an effective pair potential, our model treats these excluded volume interactions explicitly through mobile spheres attached to the micelle surface.  This allows us to (1) study the self-assembly of the micelles and (2) directly measure the relative effect of entropy and energy in driving the stabilization of assembled phases.  The MSM consists of a non-interacting rigid scaffolding with 42 points on the surface of a sphere, given by the vertex points of a 2-frequency icosahedral geodesic with diameter = $5.27\sigma$.  With this diameter, the average spacing between surface points is $1.5\sigma$.  Each surface point anchors a small spherical particle with diameter $\sigma$.  The particles and surface points are attached by harmonic springs of stiffness $k$ that control the degree of surface particle mobility.  Surface particle mobility increases as $k$ decreases, creating a larger, ``squishier'' outer corona.  Decreasing $k$ can also be interpreted as increasing the radius of gyration of the surface coating, if we consider the spheres to be dumbbell polymers anchored to the surface~\cite{larson}. Excluded volume interactions between the surface spheres are modeled by the purely-repulsive Weeks-Chandler-Andersen (WCA) potential~\cite{wca} (see Materials).  Roughly speaking, the MSM can represent many different nanoscopic objects, including core-satellite nanoparticles~\cite{mucic98,lee05,sebba08}, where nanospheres are functionalized with an outer coating of smaller nanospheres; spherical micelles composed of dendrimers~\cite{kamien,kamien00,zeng04} where the outermost layer of the dendrimer ``tree'' is functionalized with oligomers or polymers;  spherical block copolymer micelles~\cite{lee10, fischer11} that possess an outer corona of polymers; or spherical micelles created from amphiphilic tethered nanoparticles~\cite{zhang03,iacovella05, iacovella09}, as we discuss later.

\begin{figure}
\includegraphics[width=.4\textwidth]{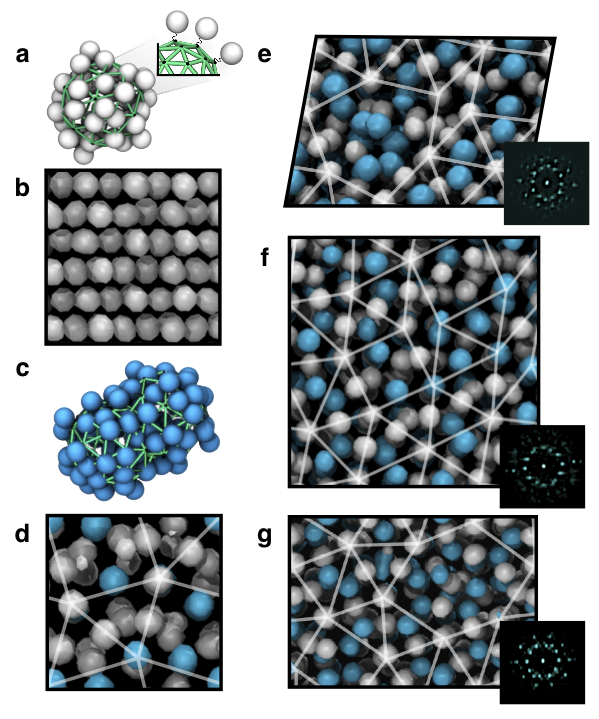}
\caption{MSM systems. (a) MSM monomer (white) extracted from a simulation. (b) 60 MSMs at $k =$ 5 with $f_{\textrm{dimer}} = 0$ (no dimers) in a bcc configuration.  (c) MSM dimer (cyan) extracted from a simulation.  (d) 60 MSMs with $k=5$ and $f_{\textrm{dimer}} \approx 0.24$ in a sigma structure.  (e-g) Systems with 360 MSMs and (e) $k=$4 and $f_{\textrm{dimer}} \approx 0.39$ , (f) $k=$4.75 and $f_{\textrm{dimer}} \approx 0.37$ , and (g) $k=$5 and $f_{\textrm{dimer}} \approx 0.36$. In all cases, we plot time-averaged density isosurfaces of the centers of mass of the micelles/dimers, rather than the micelles themselves, to remove thermal noise and produce a clearer image (see Materials).  Systems are viewed along the pseudo-12-fold-symmetry axis, as calculated using the diffraction pattern, shown to the right of each panel.  In all cases, systems are colored-coded with monomer centroids shown in white/grey and dimer centroids shown in cyan. Note, (e) appears as a parallelogram due to the projection; all simulation boxes are square or rectangular cuboids.}\label{fig:SA}
\end{figure}

In the absence of shape polydispersity, the MSMs tend to form close-packed (face centered cubic, fcc, or hexagonally close packed, hcp) arrangements for $k>5$ (lower surface particle mobility) and bcc structures for $k \leq 5$ (higher surface particle mobility); structures are identified using the algorithms described in Ref.~\cite{arcmp}.   These results support the conjecture that increasing surface particle mobility drives the system towards structures with lower surface contact area (such as bcc), as we discuss in detail in the following section.  At these statepoints, sphere packing constraints favor the bcc structure over the surface-contact-area-minimizing A15 structure.  A bcc-ordered structure of 60 MSMs is shown in Fig.~2b for $k$=5.  

We find a more dramatic change in the structural arrangement of the MSMs when shape polydispersity is incorporated into the system in the form of aspherical ``dimer'' micelles (see Fig.~2c).  We allow dimers to form in an unbiased manner by exploiting the fact that at low $k$, surface particles are only loosely bound to the surface sites on the scaffold, allowing the MSMs to overlap; some of the MSMs become locked together into dimers when $k$ is increased.  By slowly increasing from a highly-disordered state at $k=2$, we create systems with dimer fraction in the range $0.20 \leq f_{\textrm{dimer}} < 0.40$, consisting of dimers with an average aspect ratio of $\sim$1.45:1.  This procedure roughly mimics the process by which micelles are formed in amphiphilic soft matter systems, such as the tethered nanoparticle models that we discuss later. In such systems, spherical micelles assemble from a disordered mixture of individual building blocks as the system temperature is reduced ~\cite{iacovella05,horsch06} (see Materials).  In the MSM system, increasing $k$ has a similar effect to decreasing the temperature.

We find that systems with a mixture of spherical and dimer MSMs consistently form FK structures~\cite{frank59}.  Fig. 2d shows a typical sigma approximant formed by 60 MSMs at $k=5$ with $f_{\textrm{dimer}} = 0.24$; sigma structures were reproducibly observed in over 25 independent simulations where $k$ was slowly increased from 2 to 5. This approximant closely matches the expected result for 60 particles interacting via the Dzugutov or Lennard-Jones-Gauss pair potentials at densities that yield DQCs for larger systems.  The formation of the sigma structure is also consistent with the observed experimental behavior of spherical dendrimer~\cite{ungar03} and block copolymer micelles~\cite{lee10}.  Three representative independent simulations, each composed of 360 MSMs in rectangular boxes with aspect ratio 1.28:1.28:1.00, are presented in Figs. 2e-g.  Figs. 2e,f, and g show systems at $k=$4, 4.75, and 5, with $f_{\textrm{dimer}} =$0.39, 0.37, and 0.36, respectively.  In all cases, we observe finite-size DQCs that exhibit long-range rotational order of the MSM center-of-mass but no periodicity aside from the trivial periodicity imposed by the boundary conditions.   Our simulations are limited to smaller system sizes than typical point-particle models~\cite{dzug93, engel08} because we must resolve timescales corresponding to the microscopic motions of the surface particles that comprise the MSM, rather than the MSM centroid.  Nevertheless, the finite structures depicted in Fig. 2e-g exhibit local indicators of DQC ordering.  The systems form unique tilings with different configurations rather than any particular approximant.  The systems also contain the entire range of valid tiles, rather than containing squares and triangles exclusively like the sigma phase, which often competes with DQCs for stability.  Since DQCs grow more easily than approximants~\cite{keys07}, it is possible that the DQC-like tilings are thermodynamically metastable relative to a stable approximant.  The structures do not rearrange or undergo phason flips after solidification during the timescale of our simulations.  

\begin{figure}
\includegraphics[width=.4\textwidth]{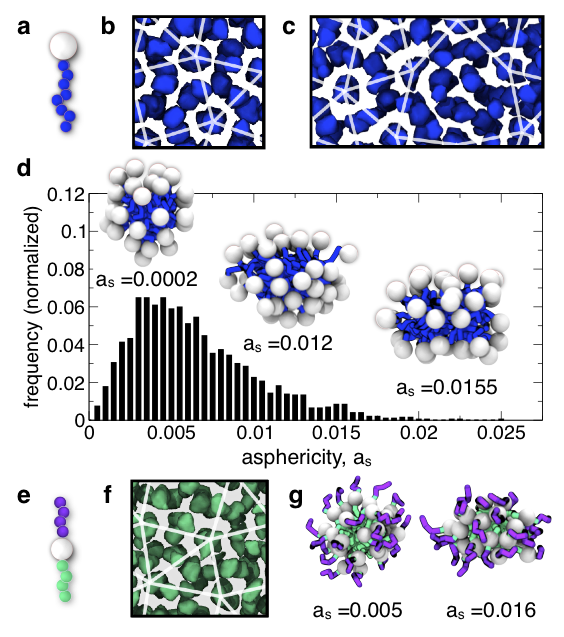}
\caption{TNS systems. (a) Schematic of a mono-TNS building block, where the 8 tether beads (blue) of size $\sigma$ aggregate and self-assemble spherical micelles with a soft core surrounded by relatively hard ``satellite'' nanoparticles (white) of size 2.5$\sigma$ that act as mobile surface entities. (b) A simulation snapshot of $\sim$60 micelles formed by mono-TNS that arrange into a sigma approximant, and (c) $\sim$120 mono-TNS micelles that form a DQC-like structure; for both systems $\phi=0.275$ and T=1.1. (d) Histogram of asphericity, $a_s$, of the mono-TNS micelles in the sigma phase.  (e) Schematic of the di-TNS building block, where the 4 beads in the tether (green) each of size $\sigma$ aggregate and nanoparticles (white) of size 2$\sigma$ are also attractive; 4 bead tethers (purple) of bead size $\sigma$ that do not aggregate coat the outside of the micelle.  (f) $\sim$60 di-TNS micelles arranged in a sigma approximant at $\phi=0.2$ and T = 1.2.  (g) Representative di-TNS micelles with different $a_s$.  In all cases, for clarity, we show density isosurfaces of the aggregating polymer tethers (i.e., the micelle core).
}\label{fig:tns}
\end{figure}
  
We can further test our proposed strategy in systems where we do not have explicit control over surface particle mobility or shape polydispersity, but where these two key features instead emerge naturally as a result of phase separation. We consider two model tethered nanosphere (TNS) systems, mono-TNS~\cite{zhang03,iacovella05} and di-TNS~\cite{zhang03,iacovella09, iacovella09b}, both of which form roughly spherical micelles with mobile surface entities.  Schematics of the building blocks are shown in Figs.~3a,e respectively, and the micelles they form are shown in Figs.~3d,g respectively.  The mono-TNS micelles have an outer shell of mobile nanospheres that closely match the MSM model, while the di-TNS micelles have a shell of short polymers, more closely resembling the spherical micelles formed by block copolymers~\cite{lee10,grason,grasonprl} and functionalized dendrimers\cite{zeng04,kamien,kamien00}.   These models are computationally expensive, and thus only relatively small systems in terms of the number of micelles are explored.  Fig.~3b depicts density isosurfaces~\cite{VMD} of the aggregating tethers for a system of 2500 mono-TNS building blocks that assemble into $\sim$60 spherical micelles arranged in a sigma approximant.  Fig.~3c depicts isosurfaces for a system of 5000 mono-TNS that self-assemble into $\sim$120 spherical micelles arranged in a FK structure containing squares, triangles, shields, and rhombs.  The increasing complexity of the tiling arrangement with system size indicates that the TNS system may form a higher-order approximant or a DQC in the infinite limit. The mono-TNS micelles naturally exhibit shape polydispersity.  Fig.~3d shows a histogram of the asphericity, $a_s$, computed from the principle radii of gyration~\cite{horsch06} of the micelles, with representative micelles at various values of $a_s$ depicted in the inset.  For reference, $a_s$ = 0 corresponds to an ideal sphere and $a_s$ = 0.02 corresponds to the MSM dimer with aspect ratio 1.45:1 shown in Fig. 2c.  Fig.~3f shows a sigma structure formed from 2000 di-TNS building blocks that self-assemble into $\sim$60 micelles. The distribution of $a_s$ for the di-TNS (plotted in the Fig. 6) is similar to that for the mono-TNS system.  Two representative di-TNS micelles at low and high $a_s$ are depicted in Fig. 3g.  Overall, FK structures self-assembled from TNS building blocks were reproducibly observed in 20 independent simulations. Whether these systems form DQCs in the infinite limit remains an open question that should be explored in the future as computational power increases.   

\section{Free-energy calculations}

Having explored the self-assembly of the three micelle models, we now perform free-energy calculations to investigate the thermodynamic basis underlying both aspects of our strategy for DQC-like structure stabilization.  The first aspect, the functionalization of particles with mobile surface entities, is inspired by the observation that soft-matter systems with relatively soft inter-micelle interactions often form non-close packed structures, as described in the Introduction.  For example, spherical dendrimeric micelles functionalized with alkyl tails to create a ``squishy corona'' are known to form non-close packed structures such as the bcc and A15 crystals~\cite{balagurusamy_jacs}.  Ziherl and Kamien proposed that the formation of the bcc and A15 structures is related to the Kelvin problem, which involves finding the space-filling arrangement of cells that minimizes surface contact area~\cite{kamien,kamien00}.  In this picture, the dendrimeric micelles adopt structures with low surface contact area in order to reduce steric interactions between terminal polymer groups.  The bcc and A15 crystals both exhibit low surface contact area, with A15 representing the current best-known solution to the Kelvin problem~\cite{weaire}.  It has been suggested~\cite{zeng04} that this mechanism may also stabilize the dendrimer DQC observed in experiments~\cite{zeng04}. However, since minimizing surface area alone favors the A15 structure rather than the DQC, other factors must be important as well.

We calculate the Helmholtz free energy, $F$~\cite{zwanzig, kofkefep}, as a function of the surface particle mobility $k$ for a system of monodisperse MSMs (i.e., without dimers); see the Materials for more information.  The value of $F$ in Fig.~4a is shown relative to the value for the hcp crystal, taken as a convenient reference state.  Fig.~4a illustrates that as $k$ decreases (i.e., surface particle mobility increases), $F$ decreases more rapidly for the A15, dod, and bcc structures than for the fcc and hcp structures.  Here, the value for the ``dod'' curve is the average of the sigma phase and several higher-order square-triangle approximants to the DQC~\cite{zeng2006inflation}, all of which have nearly identical free energies.  For low $k$, bcc appears to be the stable state, consistent with our MSM simulation results.  For very low $k$ ($k < 3$) the system becomes disordered.  The change in $F$ as a function of $k$ is the strongest for the A15 structure, which minimizes surface contact area, followed by the dod and bcc structures, respectively. We note that the dod structure has a lower free energy than the A15 structure over the entire range; however, at sufficiently low $k$, the difference in free energy between bcc, A15, and dod is indistinguishable.  The change in $F$ with $k$ is entropically driven; the difference in average potential energy $\left< U \right>$ changes little, and does not decrease with $F$ (Fig. 4a, inset).  This serves as a direct verification of the predictions of Ziherl and Kamien~\cite{kamien,kamien00}.  Note that the Z structure (Fig.~1c) is omitted as it is not stable in the parameter range under consideration.  While the trends in entropy are as we expect, we find surface particle mobility alone is not sufficient to stabilize DQCs or approximants for the statepoints and model under consideration. Thus, as our self-assembly simulations previously showed, a second mechanism is needed to form DQC structures for this model.

\begin{figure}
\includegraphics[width=.43\textwidth]{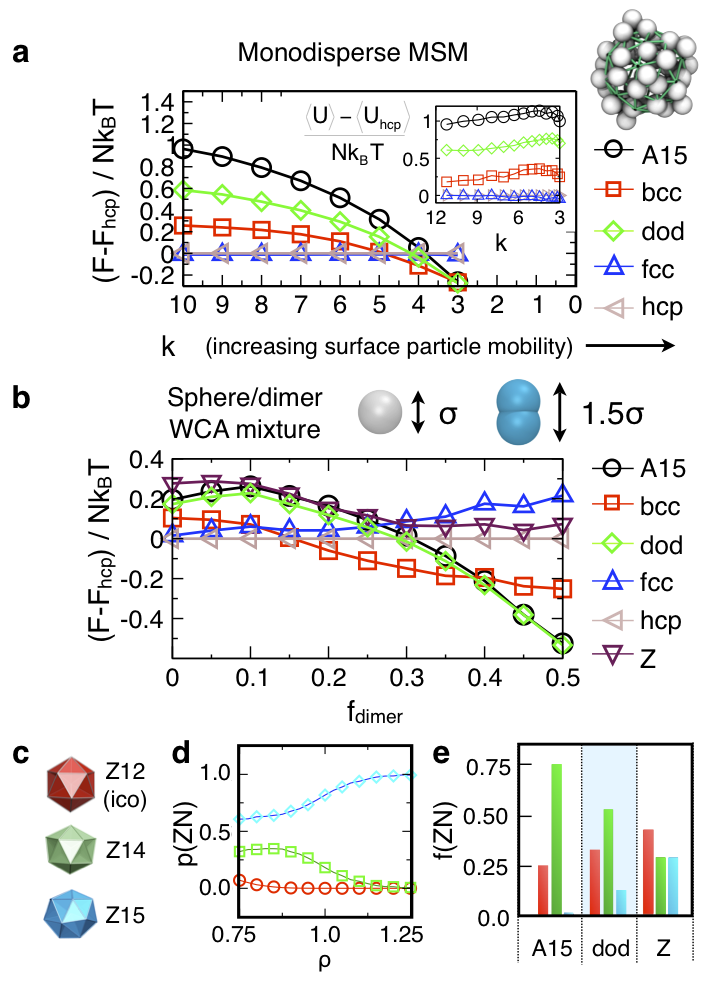}
\caption{(a) Helmholtz free energy per micelle, $F$, as a function of surface particle mobility (controlled by the spring stiffness $k$) for monodisperse MSMs.  The inset shows the potential energy per micelle $U$.  (b) $F$ as a function of $f_{\textrm{dimer}}$ for the Weeks-Chandler-Andersen sphere/dimer mixture.  For parts (a) and (b), the energies and free energies are reported with respect to the hcp crystal for convenience.  For all datapoints, error bars are smaller than the data labels. (c) Depiction of three different FK polyhedra. (d) Probability of observing dimers at the center of Z12, Z14, and Z15 structures in the dod phase as a function of number density, $\rho$. (e) Fraction of Z12, Z14, and Z15 local structures in A15, sigma and Z structures.  Note, (d-e) are color coded following the convention in (c). 
}\label{fig:FE}
\end{figure}

The thermodynamic basis underlying the second aspect of our strategy -- shape polydispersity -- can be understood in the context of previous studies of both quasicrystal formation and sphere packing.  Systems of particles with short-ranged, spherically-symmetric interaction potentials, such as hard spheres or particles with short-ranged van der Waals interactions, modeled by the Lennard-Jones (LJ) potential, tend to form close-packed crystals in the solid phase, e.g. f and/or hcp.  Although these systems tend to locally favor polytetrahedral structures~\cite{frank58}, close-packed structures maximize the overall packing density and hence maximize the entropy, and also often exhibit low potential energy.  Specialized interparticle potentials, such as the Dzugutov~\cite{dzug93} and Lennard-Jones-Gauss~\cite{engel08} potentials, have been contrived with features that help drive systems away from close-packed structures.  Like the standard LJ potential, the Dzugutov and Lennard-Jones-Gauss potentials have an attractive well that encourages local polytetrahedral ordering. However, these specialized potentials include an additional relative energy penalty for adopting the characteristic interatomic spacings associated with close packing, ultimately driving the system to form alternative structures, such as bcc crystals, as well as DQCs and their approximants under certain conditions~\cite{roth00, engel08}.  We propose, as our previous MSM simulations show, that shape polydispersity can have a similar effect, driving the system away from close-packing.  However, in contrast to the energetic repulsion of the Dzugutov potential, the destabilizing effect, in this case, is entropic.

To explicitly quantify the effect of shape polydispersity, we perform free energy calculations~\cite{meijer90, frenkelsmit, frenkelladd, engel11, shirts2007alchemical} for binary mixtures of soft spheres and short, pill-shaped dimers, with particle interactions modeled by the WCA potential (see Materials).  The dimers are modeled by a rigid body of length $1.5\sigma$ consisting of two overlapping soft spheres $0.5\sigma$ apart (see Fig.~4b), resulting in an aspect ratio of 1.5:1, similar to the aspect ratio observed in the simulation of MSMs.   Fig.~4b shows the Helmholtz free energy, $F$, as a function of the dimer fraction, $f_{\textrm{dimer}}$, for several structures at a representative state point with number density $\rho =0.9$ and $T = 0.25$.  The free energy is computed based on the standard Einstein crystal thermodynamic integration (TI) method~\cite{frenkelladd, engel11}, with an additional alchemica~l\cite{shirts2007alchemical} TI step to compute the free energy required to transform a given fraction of spheres into dimers (see Materials).   As $f_{\textrm{dimer}}$ increases, the A15 and dod structures become increasingly stable relative to close packed crystals, and, to a lesser extent, the bcc crystal.  We note that the dod phase has a lower value of $F$ than the A15 structure for all statepoints, although the difference becomes minimal for high dimer fraction.  

This difference in stability between the FK phases (A15 and dod) and standard crystals can be traced to the tendency for dimers to adopt larger, more aspherical neighbor shells, which are present in FK structures but not fcc, hcp or bcc crystals.  The first neighbor shells of particles in FK structures form different types of polyhedra, which may be icosahedral (coordination number 12), or take on higher coordination numbers Z, such as Z13, Z14 or Z15 depicted in Fig.~4c.  In Fig.~4d, we plot the probability of observing dimers in Z12, Z14, and Z15 configurations for the dod structure where we fix particle centroids but allow dimers to rotate and swap positions with monomers.  We observe that dimers strongly favor Z15 coordination shells as these are the largest and thus most accommodating.  Dimers sit in Z14 arrangements as a second best option and almost never occupy Z12 structures which are the smallest.  We can gain additional insight by examining the relative fraction of Z12, Z14, and Z15 within the three approximant structures, as shown in Fig.~4e.  Although the free energy of the A15 and dod phases are similar, the A15 phase does not possess any Z15 arrangements, whereas the dod phase has an appreciable fraction ($\sim$0.13).  This difference may account for the widespread formation of dod rather than the A15 structures in our three simulation models.  We note that while the Z phase has the largest fraction of Z15 coordinations, it also possesses the largest fraction of the less favorable Z12 coordinations, which may partially account for its relative instability for this density and dimer size.  

We observe that for $f_{\textrm{dimer}} > 0.4$, A15 and dod are more stable than fcc, hcp and bcc crystals.  This implies that mixtures of spherical and pill-shaped colloids might produce DQCs or approximants.  However, since many dimers are required to destabilize crystal structures, in practice, these mixtures may remain liquid-like, phase separate, or form other ordered structures not considered here.  Along this same line, it is possible that, in specific cases, systems may form DQCs or other FK structures based on mobile surface particles alone; the entropic effect may be stronger for terminal groups that are longer or more complex than the one-bead model tested here; however, it seems likely that the A15 structure would still demonstrate the strongest entropic response due to the minimal surface area mechanism~\cite{kamien,kamien00, weaire}.  Since asphericity is common in many micellar systems that also have soft coronas, such as the previously discussed TNS micelles, it may not be possible to completely separate these two aspects. Our results suggest that even moderate levels of asphericity may enhance the relative stability and/or range of stability of DQCs and approximants for systems with squishy surface coatings. 

\section{Conclusions}
Our results demonstrate a two-part, experimentally-feasible assembly strategy for forming 3d DQCs and their approximants that can potentially be realized for a wide variety of systems.  We have introduced three new models that form DQCs and/or approximants, including a simplified model of a spherical micelle and two tethered nanoparticle models that resemble micelle-forming systems of dendrimers~\cite{zeng04,ungar03} and block copolymers ~\cite{grason,lee10,fischer11}.  Our study lends strong numerical evidence in support of the explanation for the stability of the A15 structure in systems of dendrimer micelles~\cite{kamien,kamien00} and its subsequent adoption to help explain the formation of the spherical dendrimer DQC~\cite{zeng04, lifshitz2007soft, barkan2010stability}.  Our results imply that shape polydispersity, in addition to surface particle mobility, is likely to play a role in stabilizing DQCs and approximants in micellar systems observed in experiment.  In the future, our assembly strategy may be employed to facilitate the design of new systems that can form DQCs at the nano- and micro-scale, including dendrimers~\cite{zeng04,ungar03, balagurusamy_jacs}, surfactants, block copolymers~\cite{grason,lee10,fischer11}, and core-satellite nanoparticles~\cite{mucic98,lee05,sebba08}.  Our results also suggest that mixtures of spheres and dimers~\cite{champion07,pine05,hosein07} might, even without surface particle mobility, stabilize DQCs or approximants under certain conditions, possibly providing a trivial design rule for forming these structures. In addition to the implications regarding DQC assembly, our results illustrate a powerful design approach for assembling structures by controlling particle shape and functionality to mimic the key features of pair potentials~\cite{glotzer2007}.  This paves the way for future studies based on mapping complex interaction potentials to packing models, which can potentially render currently unrealizable systems experimentally feasible, or expand the breadth of unique structures to more general classes of systems.  

\section{Materials}
\subsection{Simulation of Model Spherical Micelles}
We perform molecular dynamics simulations in the canonical ensemble (constant number of particles, volume and temperature) of the model spherical micelles (MSMs) using the LAMMPS molecular simulation package~\cite{lammps} with periodic boundary conditions and reduced Lennard-Jones (LJ) units~\cite{frenkelsmit}.  The Nose-Hoover thermostat is used with T=1.0 and timestep = 0.005. The simulations that we report in this work are performed at a \textit{nominal} volume fraction, $\phi \approx$ 0.54, computed assuming micelles are space-excluding spheres of diameter 5.27$\sigma$.  $\phi \approx$ 0.54 is chosen because it is comparable to the volume fraction of statepoints observed to form spherical micelles in previous simulations of tethered nanospheres~\cite{iacovella05}.  A cursory sensitivity analysis, performed by starting from an assembled sigma phase at $k$=5 and either slowing increasing or slowly decreasing the box size predicts that the sigma phase is physically stable within a range of \textit{nominal} volume fractions 0.47 $ \le \phi  \le $0.57, for the specific MSM parameters used in this study.  In practice, the \textit{effective} micelle volume fraction will be lower than the \textit{nonimal} volume fraction, since the micelles have a bumpy, soft corona and will overlap to form dimers.  For example, if we account for dimers, the volume fraction is reduced from 0.54 to 0.46 for a system with $f_{\textrm{dimer}}=0.25$, assuming dimers are sphero-cylinders of length 6.72 $\sigma$ and diameter 5.27$\sigma$.  In Fig. 5 we provide a schematic of the model spherical micelle (MSM), depicting the approximate micelle corona and core sizes, as well as the dimer to monomer size ratios, for the MSM simulations investigated in this paper.  

As discussed in the main text, surface particles interact via the purely-repulsive Weeks-Chandler-Anderson (WCA) potential, meant to capture excluded volume.  The WCA potential follows the form~\cite{wca}:
\begin{equation}
U = 
\begin{cases}
4 \epsilon \left( \dfrac{\sigma^{12}}{(r-\alpha)^{12}}-\dfrac{\sigma^{6}}{(r-\alpha)^6} \right) - U_{s}\;, & (r-\alpha)<r_{c} \\
0\;, &  (r-\alpha)\ge r_{c}
\end{cases}
\label{eqnLJWCA}
\end{equation}

\noindent
where $U_{s} = -\epsilon$ and the interaction cutoff $r_{c}$, is set to $2^{1/6}$. $\alpha$ is an adjustable shifting parameter, set to zero here. Surface particles are held to their scaffold sites with harmonic springs, defined as $U(r)=kr^2$, following the convention in LAMMPS~\cite{lammps}. The general simulation procedure is as follows: simulations are initialized by creating a random arrangement of MSMs under dilute conditions ($\phi <$ 0.1) with $k$=10; the system is then slowly compressed until the target box size is reached.  Starting from the target box size, MSMs are simulated with $k$=2 and further allowed to disorder.  $k$ is then incrementally increased by 0.25 until the final value is reached (typically, $k$ = 4 to 5).  For each value of $k$, the system is run for 10 to 50 million timesteps (large systems are run for longer than small systems).  Simulations are typically run for between 50 and 500 million total timesteps.  This procedure of slowly increasing $k$ mimics the procedure used to simulate tethered nanospheres, discussed below.  Note, our simulations did not span long enough time scales to observe tile rearrangements in the ordered solid phases.

\begin{figure}[ht]
\begin{center}
\includegraphics[width=.4\textwidth]{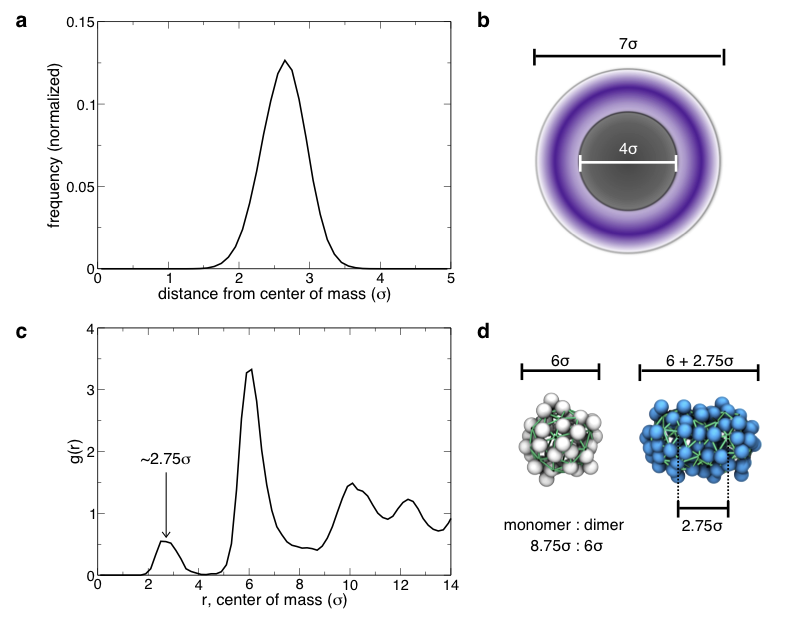}
\caption{\textbf{(a)} Histogram of the radial distance between the MSM centroid and its mobile surface spheres at $k$=5. \textbf{(b)} Based on (a), a schematic of the relative size ratio of the corona to micelle core in the MSM, where we find the micelle core diameter is roughly 4$\sigma$ and the corona extends to a diameter of 7$\sigma$.  \textbf{(c)} The radial distribution function, g(r), calculated between the centroids of MSMs arranged in a sigma structure at $k$=5. Note the first peak at $\sim$2.75$\sigma$ corresponds to the center-to-center distance of two overlapping micelles (i.e., a dimer). The second peak at $\sim$6$\sigma$ corresponds to the spacing between non-overlapping micelles. \textbf{(d)} Schematic of the monomer and dimer size ratios, as derived from g(r) in (c).  All systems at nominal volume fraction $\phi = 0.54$.}
\label{size}
\end{center}
\end{figure}

\subsection{Simulation of Tethered Nanospheres}
We perform Brownian dynamics simulations in the canonical ensemble of the TNS system with periodic boundary conditions and reduced LJ units.  The mono-TNS building block used consists of a chain of 8 spherical beads of diameter $\sigma$, connected via finitely extensible non-linear elastic (FENE) springs\cite{zhang03, iacovella05}.  Each chain is attached with a FENE spring to a larger ``nanosphere'' of diameter D = 2.5$\sigma$.  The potential energy of the FENE spring is given by
\begin{equation}
U_\textrm{FENE}(r) = -\frac{1}{2}kR_o^2 \ln \left[1-\left( \frac{(r-\alpha)}{R_o}\right)^2\right],
\label{eqnFENE}
\end{equation}
\noindent
where $k$ is the spring constant, $r$ is the interparticle separation, $R_o$ is the maximum allowable separation, and $\alpha$ is an adjustable shifting parameter to account for excluded volume of the nanosphere.  For this study, $k$=30 and $R_o$= 1.5, and $\alpha$ = 0.75 for the bond connecting the chain to the nanosphere, and zero otherwise.  Tethers are treated as ``solvent-phobic'' and thus aggregate at sufficiently low T. To model this aggregation, the attractive LJ potential is used, give by Eqn. \ref{eqnLJWCA}, but with $U_{s}$ set to the energy at the cutoff and $r_{c} = 2.5$.   All other interactions are modeled by the purely repulsive WCA potential (Eqn. \ref{eqnLJWCA}, with $U_{s}=-\epsilon$ and $r_{c} = 2^{1/6}$), appropriately radially-shifted to account for excluded volume; for tether-nanopshere interactions $\alpha = 0.75\sigma$ and for nanosphere-nanosphere interactions $\alpha = 1.5\sigma$.   Simulations are performed using the Brownian dynamics thermostat, where the volume fraction of the excluded volume of the individual beads is varied between 0.25 $\le \phi \le$ 0.30, the range where spherical micelles were predicted in previous work \cite{iacovella05}.  This translates to a nominal micelle volume fraction of $\approx$0.52, calculated assuming a characteristic diameter of 12$\sigma$ for the spherical micelles (approximated from the radial distribution function for micelle centers).  

Di-TNS are modeled in much the same way as mono-TNS described above. Chains composed of 4 beads of diameter $\sigma$ are connected via FENE springs (Eqn. \ref{eqnFENE}). Two chains are connected to a single nanosphere of diameter D = 2.0$\sigma$, diametrically opposed. This planar angle of 180 degrees between the chains is maintained by the use of a harmonic spring between the first beads of the two polymers, with $k$=30 and equilibrium separation set to 3$\sigma$. The two polymer chains are chemically distinct. One chain is considered to be solvent-phobic (i.e., attractive), and thus is modeled by the LJ potential. The other chain is considered to be solvent-philic (i.e., non-attractive) and modeled by the WCA potential. Nanosphere-nanosphere interactions are modeled with the LJ potential, appropriately radially-shifted to account for excluded volume ($\alpha$ = 1.0). All other interactions are modeled by the WCA potential, appropriately radially-shifted ($\alpha$ = 0.5 for tether-nanosphere interactions).  Simulations are performed at $\phi$ = 0.20, as calculated from the excluded volume of the individual beads.   In Fig. 6 we plot the asphericity histogram of the micelles formed by the di-tethered nanosphere system.  We note the asphericity histogram closesly matches the result obtained for the mono-TNS system shown in Fig. 3d of the main text.

\begin{figure}
\begin{center}
\includegraphics[width=.4\textwidth]{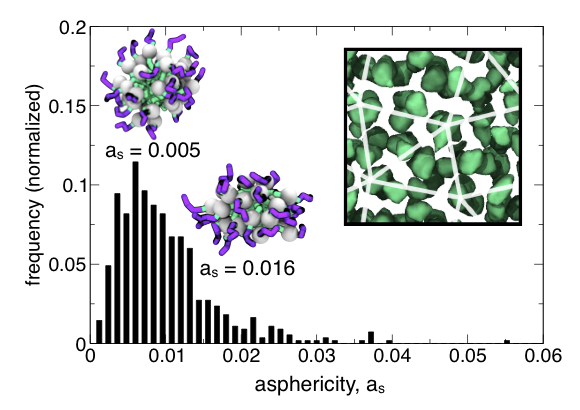}
\caption{Asphericity histogram of the sigma phase formed by the di-tethered nanosphere system.  Two representative micelles are inset in the plot along with an image of the sigma phase formed by the di-tethered nanospheres.}
\label{dtns}
\end{center}
\end{figure}

The general simulation procedure used is identical for both mono- and di-TNS.  Systems start from a disordered mixture of TNS, well above the order-disorder temperature where little-to-no aggregation occurs (T=2.0).  Systems are then incrementally cooled to their final temperature (T$\approx$1.0), where systems are run for several million timesteps at each incremental temperature. The potential energy is monitored to ensure a steady state is reached before additional cooling. As the temperature is reduced, individual TNS slowly aggregate into micelles; micelles form ordered structures at sufficiently low T.  Simulations are typically run for a total of 40 million timesteps. Multiple independent cooling sequences are performed to ensure reproducibility of results.  Simulations of mono-TNS are performed for systems of 2500 building blocks in cubic boxes (22500 total beads) and 5000 building blocks in boxes with aspect ratio 2:2:1 (45000 total beads).  Simulations of Di-TNS are performed in cubic boxes for systems of 2000 building blocks. See References \cite{zhang03,iacovella05, iacovella09, iacovella09b} for additional details regarding the simulation of TNS. Note, our simulations did not span long enough time scales to observe tile rearrangements in the ordered solid phases.

\subsection{Free Energy of Model Spherical Micelles}
We investigate how surface particle mobility affects the stability of MSMs in various crystal structures (Fig. 4a) by calculating the change in Helmholtz free energy $F$ as a function of the strength of the harmonic springs (i.e., $k$) that anchor particles to the micelle surface, using free energy perturbation (FEP)~\cite{zwanzig}. For each structure, the calculation is split into eight independent stages to avoid asymmetric bias\cite{kofkefep}, consisting of equilibrium simulations with spring constants $ k=3,4... 10$.  Within each stage, we compute the ensemble average:
\begin{equation}
F(k_j) - F(k_i) = -k_B T \ln \left \langle \exp \left ( - \frac{E_{k_j} - E_{k_i}}{k_B T} \right ) \right \rangle _{k_i} ,
\end{equation}
which gives the free energy difference between stage $i$ and $j$, where $j = i \pm 1$.  Each stage consists of a molecular dynamics simulation~\cite{lammps} in the canonical ensemble at $T=1.0$ and micelle nominal volume fraction $\phi \approx$ 0.54 to match the simulations.   The vertical offset between the curves for different structures is computed using a FEP variant of the standard Frenkel-Ladd lattice coupling expansion thermodynamic integration (TI) method for molecular systems~\cite{meijer90, frenkelsmit}.  We use FEP to adapt the method to the complex objects considered here.  Although this method is non-standard, it gives reasonable estimates of $F$ that are consistent with our self-assembly results. We note this calculation only effects the vertical offset of the curves and not how $F$ changes as a function of $k$.  

\subsection{Free Energy of Sphere/Dimer Mixtures}

The free energy for mixtures of WCA spheres and dimers is evaluated using a three-step scheme based on the standard Einstein crystal TI method for spherical particles~\cite{frenkelladd, frenkelsmit, engel11}, plus an additional alchemical~\cite{shirts2007alchemical} step to convert a given fraction of the spheres into dimers.  Since systems of WCA spheres do not act like harmonic crystals for many of the structures tested, we use the Dzugutov (DZ)~\cite{dzug93} system as a convenient reference system that gives harmonic behavior.  Computing the work required to change the particle interactions from the DZ potential to the WCA potential constitutes the third step of our scheme.

In the first step, we compute the free energy difference between a non-interacting harmonic (Einstein) crystal and a system of spherical particles interacting with the DZ potential using the standard Frenkel-Ladd method\cite{frenkelladd, frenkelsmit, engel11}.  We denote this free energy difference $\Delta F_\textrm{I} = F_\textrm{DZ} - F_\textrm{Ein}$.

In the second step, we compute the work required to change a given fraction, $f_d$ of the spherical particles in the system into dimers.  We consider a system with the energy function:
\begin{equation}
U(\lambda)_\textrm{II} = (1-\lambda)U_\textrm{pure} + \lambda U_\textrm{mix}.
\end{equation}
The free energy required to change the system of DZ spheres (pure) to a mixture of DZ spheres and dimers (mix) is the integral over the derivative with respect to the so-called switching parameter $\lambda$:
\begin{equation}
\Delta F_\textrm{II} =  \int_{0}^{1} d \lambda \left< \frac{ \delta U(\lambda) }{ \delta \lambda} \right>_\lambda 
=  \left<  U_\textrm{mix} - U_\textrm{pure} \right>_\lambda.
\end{equation}
The third step is to compute the work required to change the potential from DZ to the WCA potential. We consider a system with the energy function:
\begin{equation}
U(\lambda)_\textrm{III} = (1-\lambda)U_\textrm{DZ} + \lambda U_\textrm{WCA}.
\end{equation}
The free energy difference for changing the interaction potential is given by integrating over the derivative with respect to the switching parameter $\lambda$:
\begin{equation}
\Delta F_\textrm{III} =  \int_{0}^{1} d \lambda \left< \frac{ \delta U(\lambda) }{ \delta \lambda} \right>_\lambda 
=  \left<  U_\textrm{WCA} - U_\textrm{DZ} \right>_\lambda.
\end{equation}
For all calculations, we run 20 independent MC simulations for different values of $\lambda$ to estimate $\delta U(\lambda) / \delta \lambda$, and obtain $\Delta F$ by numerical integration.  Simulations are selectively carried out where $| \delta^2 U(\lambda) / \delta \lambda^2 |$ is the largest.  For systems that contain dimers, each simulation begins with a compression run with particles constrained to their lattice positions allowing rotations and swaps before equilibrating at constant density.  The total free energy for a sphere/dimer mixture is given by:
\begin{equation}
F = F_\textrm{Ein} + \Delta F_\textrm{I} + \Delta F_\textrm{II} + \Delta F_\textrm{III}.
\end{equation}
This formula is used to evaluate $F$ for the WCA system, shown in Fig. 4b.

\subsection{Isosurface Generation}

Representative configurations from self-assembly simulations are plotted as time-averaged isosurfaces to coarse-grain over thermal fluctuations and produce a clearer picture of the structure.  To generate the isosurfaces, we replace the centroids of the MSMs (or aggregating tethers in the case of TNS) with a normalized Gaussian of width 1.5$\sigma$, mapped to a voxel grid composed of cells of length 1$\sigma$, to achieve a degree of spatial coarse-graining.  We then average the voxel data for ten configurations generated within a time window that is much shorter than diffusion timescales.  This voxel data is then used to create isosurfaces within the Visual Molecular Dynamics software program ~\cite{VMD}, with an isovalue typically ranging between 0.1 and 0.2 for MSMs and between 4 and 6 for the TNS systems.

\section{acknowledgments}
We thank Michael Engel and Ron Lifshitz for helpful discussions.  Simulations of the model spherical micelle and tethered nanoparticle systems were supported by the U.S. Department of Energy, Office of Basic Energy Sciences, Division of Materials Sciences and Engineering under Award \# DE- FG02-02ER46000 (C.R.I. and S.C.G.). The free energy studies were supported by the National Science Foundation, Division of Chemistry, under Award \# CHE 0624807 (A.S.K. and S.C.G.).  S.C.G. is supported by a National Security Science and Engineering Faculty Fellow Award under the auspices of the Department of Defense/Director, Defense Research and Engineering. This material is based upon work supported by the DOD/DDRE under Award No. N00244-09-1-0062. Any opinions, findings, and conclusions or recommendations expressed in this publication are those of the author(s) and do not necessarily reflect the views of the DOD/DDRE.


\begin{thebibliography}{10}

\bibitem{grier}
Roichman, Y \& Grier, D.~G.
\newblock (2005) Holographic assembly of quasicrystalline photonic
  heterostructures {\em Opt. Express} {\bf 13}, 5434--5439.

\bibitem{mikhael}
Mikhael, J, Roth, J, Helden, L,  \& Bechinger, C.
\newblock (2008) Archimedean-like tiling on decagonal quasicrystalline surfaces
  {\em Nature} {\bf 454}, 501--504.

\bibitem{glotzer08}
Glotzer, S.~C \& Keys, A.~S.
\newblock (2008) Materials science: A tale of two tilings {\em Nature} {\bf
  454}, 420--421.

\bibitem{ungar03}
Ungar, G, Liu, Y, Zeng, X, Percec, V,  \& Cho, W.~D.
\newblock (2003) Giant supramolecular liquid crystal lattice {\em Science} {\bf
  299}, 1208--1211.

\bibitem{zeng04}
Zeng, X, Ungar, G, Liu, Y, Percec, V, Dulcey, A.~E,  \& Hobbs, J.~K.
\newblock (2004) Supramolecular dendritic liquid quasicrystals {\em Nature}
  {\bf 428}, 157--160.

\bibitem{dotera}
Hayashida, K, Dotera, T, Takano, A,  \& Matsushita, Y.
\newblock (2007) Polymeric quasicrystal: Mesoscopic quasicrystalline tiling in
  abc star polymers {\em Phys. Rev. Lett.} {\bf 98}, 195502.

\bibitem{talapin2009}
Talapin, D.~V, Chevchenko, E.~V, Bodnarchuk, M.~I, X., Y, Chen, J,  \& Murray,
  C.~B.
\newblock (2009) Quasicrystalline order in self-assembled binary nanoparticle
  superlattices {\em Nature} {\bf 461}, 964--967.

\bibitem{lee10}
Lee, S, Bluemle, M.~J,  \& Bates, F.~S.
\newblock (2010) Discovery of a frank-kasper {sigma} phase in sphere-forming
  block copolymer melts {\em Science} {\bf 330}, 349--353.

\bibitem{fischer11}
Fischer, S, Exner, A, Zielske, K, Perlich, J, Deloudi, S, Steurer, W, Lindner,
  P,  \& F{\\"o}rster, S.
\newblock (2011) Colloidal quasicrystals with 12-fold and 18-fold diffraction
  symmetry {\em Proc. Natl. Acad. Sci.} {\bf 108}, 1810.

\bibitem{amir09}
Haji-Akbari, A, Engel, M, Keys, A.~S, Zheng, X, Petschek, R.~G, Palffy-Muhoray,
  P,  \& Glotzer, S.~C.
\newblock (2009) Disordered, quasicrystalline and crystalline phases of densely
  packed tetrahedra {\em Nature} {\bf 462}, 773--777.

\bibitem{balagurusamy_jacs}
Balagurusamy, V. S.~K, Ungar, G, Percec, V,  \& Johansson, G.
\newblock (1997) Rational design of the first spherical supramolecular
  dendrimers self-organized in a novel thermotropic cubic liquid-crystalline
  phase and the determination of their shape by x-ray analysis {\em J. Am.
  Chem. Soc.} {\bf 119}, 1539--1555.

\bibitem{kamien}
Ziherl, P \& Kamien, R.~D.
\newblock (2001) Maximizing entropy by minimizing area: Towards a new principle
  of self-organization {\em Phys. Chem. B} {\bf 105}, 10147--10158.

\bibitem{kamien00}
Ziherl, P \& Kamien, R.
\newblock (2000) Soap froths and crystal structures {\em Phy. Rev. Lett.} {\bf
  85}, 3528--3531.

\bibitem{weaire}
Weaire, D \& Phelan, R.
\newblock (1994) A counterexample to kelvin's conjecture on minimal surfaces
  {\em Phil. Mag. Lett.} {\bf 69}, 107--110.

\bibitem{lifshitz2007soft}
Lifshitz, R \& Diamant, H.
\newblock (2007) Soft quasicrystals-why are they stable? {\em Philos. Mag.}
  {\bf 87}, 3021--3030.

\bibitem{barkan2010stability}
Barkan, K, Diamant, H,  \& Lifshitz, R.
\newblock (2011) Stability of quasicrystals composed of soft isotropic
  particles {\em Phys. Rev. B} {\bf 83}, 172201.

\bibitem{dzug93}
Dzugutov, M.
\newblock (1993) Formation of a dodecagonal quasicrystalline phase in a simple
  monatomic liquid {\em Phys. Rev. Lett.} {\bf 70}, 2924--2927.

\bibitem{roth00}
Roth, J \& Denton, A.~R.
\newblock (2000) Solid-phase structures of the dzugutov pair potential {\em
  Phys. Rev. E} {\bf 61}, 6845--6857.

\bibitem{engel08}
Engel, M \& Trebin, H.~R.
\newblock (2008) Structural complexity in monodisperse systems of isotropic
  particles {\em Z. Kristallogr.} {\bf 223}, 721--725.

\bibitem{glotzer2007}
Glotzer, S.~C \& Solomon, M.~J.
\newblock (2007) Anisotropy of building blocks and their assembly into complex
  structures {\em Nat. Mater.} {\bf 6}, 557--562.

\bibitem{zhang03}
Zhang, Z.-L, Horsch, M.~A, Lamm, M.~H,  \& Glotzer, S.~C.
\newblock (2003) Tethered nano building blocks: Toward a conceptual framework
  for nanoparticle self-assembly {\em Nano Lett.} {\bf 3}, 1341--1346.

\bibitem{iacovella05}
Iacovella, C.~R, Horsch, M.~A, Zhang, Z,  \& Glotzer, S.~C.
\newblock (2005) Phase diagrams of self-assembled mono-tethered nanospheres
  from molecular simulation and comparison to surfactants {\em Langmuir} {\bf
  21}, 9488--9494.

\bibitem{iacovella09}
Iacovella, C.~R \& Glotzer, S.~C.
\newblock (2009) Complex crystal structures formed by the self assembly of
  di-tethered nanospheres {\em Nano Lett.} {\bf 9}, 1206--1211.

\bibitem{iacovella09b}
Iacovella, C.~R \& Glotzer, S.~C.
\newblock (2009) Phase behavior of ditethered nanospheres {\em Soft Matter}
  {\bf 5}, 4492--4498.

\bibitem{roth00epjb}
Roth, J.
\newblock (2000) The fluid-solid transition of dzugutov's potential {\em Eur.
  Phys. J. B} {\bf 14}, 449--458.

\bibitem{lifshitz2007crystal}
Lifshitz, R.
\newblock (2007) What is a crystal? {\em Zeitschrift f{\\"u}r Kristallographie}
  {\bf 222}, 313--317.

\bibitem{lifshitz2003quasicrystals}
Lifshitz, R.
\newblock (2003) Quasicrystals: A matter of definition {\em Foundations of
  Physics} {\bf 33}, 1703--1711.

\bibitem{nelson1989polytetrahedral}
Nelson, D \& Spaepen, F.
\newblock (1989) Polytetrahedral order in condensed matter {\em Solid State
  Phys.} {\bf 42}, 1--90.

\bibitem{frank59}
Frank, F.~C \& Kasper, J.~S.
\newblock (1959) Complex alloy structures regarded as sphere packings. ii.
  analysis and classification of representative structures {\em Acta
  Crystallogr.} {\bf 12}, 483--499.

\bibitem{goldman93}
Goldman, A.~I \& Kelton, R.~F.
\newblock (1993) Quasicrystals and crystalline approximants {\em Rev. Mod.
  Phys.} {\bf 65}, 213--230.

\bibitem{janot97}
Janot, C.
\newblock (1997) {\em Quasicrystals: a primer}.
\newblock (Oxford University Press, USA).

\bibitem{zeng2006inflation}
Zeng, X \& Ungar, G.
\newblock (2006) Inflation rules of square-triangle tilings: from approximants
  to dodecagonal liquid quasicrystals {\em Philos. Mag.} {\bf 86}, 1093--1103.

\bibitem{onoda1988growing}
Onoda, G.~Y, Steinhardt, P.~J, DiVincenzo, D.~P,  \& Socolar, J. E.~S.
\newblock (1988) Growing perfect quasicrystals {\em Phys. Rev. Lett.} {\bf 60},
  2653--2656.

\bibitem{steinhardt1996simpler}
Steinhardt, P.~J \& Jeong, H.~C.
\newblock (1996) A simpler approach to penrose tiling with implications for
  quasicrystal formation {\em Nature} {\bf 382}, 431--433.

\bibitem{henley1991random}
Henley, C.~L.
\newblock (1991) Random tiling models {\em {Quasicrystals: The state of the
  art}} pp. 429--524.

\bibitem{lubensky1986distortion}
Lubensky, T.~C, Socolar, J. E.~S, Steinhardt, P.~J, Bancel, P.~A,  \& Heiney,
  P.~A.
\newblock (1986) Distortion and peak broadening in quasicrystal diffraction
  patterns {\em Phys. Rev. Lett.} {\bf 57}, 1440--1443.

\bibitem{keys07}
Keys, A.~S \& Glotzer, S.~C.
\newblock (2007) How do quasicrystals grow? {\em Phys. Rev. Lett.} {\bf 99},
  235503.

\bibitem{lammps}
Plimpton, S.~J.
\newblock (1995) Fast parallel algorithms for short-range molecular dynamics
  {\em J. Comput. Phys.} {\bf 117}.

\bibitem{larson}
Larson, R.
\newblock (1998) {\em The Structure and Rheology of Complex Fluids}.
\newblock (Oxford University Press, USA).

\bibitem{wca}
Weeks, J, Chandler, D,  \& Andersen, H.
\newblock (1971) Role of repulsive forces in determining the equilibrium
  structure of simple liquids {\em J. Chem. Phys.} {\bf 54}, 5237.

\bibitem{mucic98}
Mucic, R.~C, Storhoff, J.~J, Mirkin, C.~A,  \& Letsinger, R.~L.
\newblock (1998) Dna-directed synthesis of binary nanoparticle network
  materials {\em J. Am. Chem. Soc.} {\bf 120}, 12674--12675.

\bibitem{lee05}
Lee, J, Govorov, A.~O,  \& Kotov, N.~A.
\newblock (2005) Nanoparticle assemblies with molecular springs: A nanoscale
  thermometer {\em Angew. Chem. Int. Edit.} {\bf 44}, 7439--7442.

\bibitem{sebba08}
Sebba, D.~S, Mock, J.~J, Smith, D.~R, LaBean, T.~H,  \& Lazarides, A.~A.
\newblock (2008) Reconfigurable core- satellite nanoassemblies as
  molecularly-driven plasmonic switches {\em Nano Lett.} {\bf 8}, 1803--1808.

\bibitem{arcmp}
Keys, A.~S, Iacovella, C.~R,  \& Glotzer, S.~C.
\newblock (2011) Characterizing structure through shape matching and
  applications to self-assembly {\em Annual Review of Condensed Matter Physics}
  {\bf 2}.

\bibitem{horsch06}
Horsch, M.~A, Zhang, Z,  \& Glotzer, S.~C.
\newblock (2006) Simulation studies of self-assembly of end-tethered nanorods
  in solution and role of rod aspect ratio and tether length {\em J. Chem.
  Phys.} {\bf 125}, 184903.

\bibitem{grason}
Grason, G.~M \& Kamien, R.~D.
\newblock (2005) Self-consistent field theory of multiply branched block
  copolymer melts {\em Phys. Rev. E} {\bf 71}, 051801.

\bibitem{grasonprl}
Grason, G.~M, DiDonna, B.~A,  \& Kamien, R.~D.
\newblock (2003) Geometric theory of diblock copolymer phases {\em Phys. Rev.
  Lett.} {\bf 91}, 58304.

\bibitem{VMD}
Humphrey, W, Dalke, A,  \& Schulten, K.
\newblock (1996) {VMD: Visual molecular dynamics} {\em J. Molec. Graphics} {\bf
  14}, 33--38.

\bibitem{zwanzig}
Zwanzig, R.
\newblock (1954) {High-Temperature Equation of State by a Perturbation Method.
  I. Nonpolar Gases} {\em J. Chem. Phys.} {\bf 22}, 1420.

\bibitem{kofkefep}
Lu, N, Singh, J,  \& Kofke, D.
\newblock (2003) Appropriate methods to combine forward and reverse free-energy
  perturbation averages {\em J. Chem. Phys.} {\bf 118}, 2977.

\bibitem{frank58}
Frank, F.~C \& Kasper, J.~S.
\newblock (1958) Complex alloy structures regarded as sphere packings. i.
  definitions and basic principles {\em Acta Crystallogr.} {\bf 11}, 184--190.

\bibitem{meijer90}
Meijer, E.~J, Frenkel, D, Lesar, R.~A,  \& Ladd, A. J.~C.
\newblock (1990) Location of melting-point at 300-k of nitrogen by monte-carlo
  simulation {\em J. Chem. Phys.} {\bf 92}, 7570--7575.

\bibitem{frenkelsmit}
Frenkel, D \& Smit, B.
\newblock (2002) {\em Understanding molecular simulation: from algorithms to
  applications}.
\newblock (Academic Pr).

\bibitem{frenkelladd}
Frenkel, D \& Ladd, A. J.~C.
\newblock (1984) New monte-carlo method to compute the free energy of arbitrary
  solids - application to the fcc and hcp phases of hard-spheres {\em J. Chem.
  Phys.} {\bf 81}, 3188--3193.

\bibitem{engel11}
Engel, M.
\newblock (2011) Entropic stabilization of tunable planar modulated
  superstructures {\em Phys. Rev. Lett.} {\bf 106}, 95504.

\bibitem{shirts2007alchemical}
Shirts, M, Mobley, D,  \& Chodera, J.
\newblock (2007) {Alchemical free energy calculations: Ready for prime time?}
  {\em Annual Reports in Computational Chemistry} {\bf 3}, 41--59.

\bibitem{champion07}
Champion, J.~A, Katare, Y.~K,  \& Mitragotri, S.
\newblock (2007) Making polymeric micro- and nanoparticles of complex shapes
  {\em Proc. Natl. Acad. Sci.} {\bf 104}, 11901--11904.

\bibitem{pine05}
Cho, Y.-S, Yi, G.-R, Lim, J.-M, Kim, S.-H, Manoharan, V.~N, Pine, D.~J,  \&
  Yang, S.-M.
\newblock (2005) Self-organization of bidisperse colloids in water droplets
  {\em J. Am. Chem. Soc.} {\bf 127}, 15968--15975.

\bibitem{hosein07}
Hosien, I.~D \& Liddel, C.~M.
\newblock (2007) Convectively assembled asymmetric dimer-based colloidal
  crystals {\em Langmuir} {\bf 23}, 10479--10485.

\end{thebibliography}
\end{document}